\documentclass[letter]{aa} 

\usepackage{graphicx}
\usepackage{txfonts}
\usepackage{lastpage}
\usepackage[pdftex, colorlinks=true, linkcolor=blue, citecolor=blue, urlcolor=blue]{hyperref}
\usepackage{nameref} 
\usepackage[dvipsnames]{xcolor}
\usepackage{ulem}
\usepackage{threeparttable}
\newcommand{\mic}{$\mu$m}

\newcommand\rev[1]{#1}

\begin{document}

   \title{Variation of the disk thickness across ice bands: A method to determine ice abundances in highly inclined protoplanetary disks}
   \titlerunning{Disk thickness variations to measure ice abundance}

   \author{L. Martinien
          \inst{1}
          \and
          G. Duch\^ene\inst{1,2}
          \and
          F. M\'enard\inst{1}
          \and
          K.R. Stapelfeldt\inst{3}
          \and
          R. Tazaki\inst{4}
          \and
          J. B. Bergner\inst{5}
          \and
          E. Dartois\inst{6}
          \and
          J. A. Noble\inst{7}
          \and
          W. E. Thompson\inst{5}
          }

    \institute{Univ. Grenoble Alpes, CNRS, IPAG, 38000 Grenoble, France \\
    \email{laurine.martinien@univ-grenoble-alpes.fr}
    \and Astronomy Department, University of California Berkeley, Berkeley CA 94720-3411, USA
    \and Jet Propulsion Laboratory, California Institute of Technology, Mail Stop 321-100, 4800 Oak Grove Drive, Pasadena, CA 91109, USA
    \and Department of Earth Science and Astronomy, The University of Tokyo, Tokyo 153-8902, Japan
    \and Department of Chemistry, University of California Berkeley, Berkeley CA 94720, USA
    \and Institut des Sciences Moléculaires d’Orsay, CNRS, Univ. Paris-Saclay, Orsay, France
    \and CNRS, Aix-Marseille Université, Laboratoire PIIM, Marseille, France
             }

   \date{}

 
  \abstract
   {The James Webb Space Telescope provides unprecedented information to study ices in protoplanetary disks. However, the saturation of ice bands in highly inclined disks hinders the measurement of ice abundances using classical spectroscopy. This is unfortunate as the presence and, more importantly, the abundance of ices play a key role in, e.g., the evolution of dust (because it modifies the sticking properties) and the composition of planetesimals and exoplanetary atmospheres.}
   {To overcome this issue and quantify the ice abundance within disks, we introduce a new method based on measuring the changes in the apparent disk thickness as a function of wavelength, which is directly and quantitatively related to the grain opacity. Specifically, we expect i) that the increased opacity within ice bands should result in a thicker disk than in the adjacent continuum, and ii) the thickness variations to be proportional to the abundance of ice.}
   {We extracted the disk thickness in model images of edge-on disks containing different abundances of water ice, as well as in James Webb Space Telescope spectral imaging of four edge-on disks.}
   {For both models and observations, the disk thickness decreases toward longer wavelengths except at the positions of ice absorption features where the thickness is enhanced across the band. In the model images, we demonstrate that this effect increases with ice abundance without any hint of saturation.
   This definitely demonstrates the presence of the ice species within each disk and confirms our expectation that this method can be applied to estimate ice abundances. }
   {Thanks to this method, it will thus be possible to constrain the abundance of ice in highly inclined disks with disk model fitting. Unlike spectroscopic analyses, this method is not subject to saturation and should therefore be more robust and applicable to all disks for which the two surfaces can be resolved. 
   }


   \maketitle
%

\section{Introduction}

Ices are assumed to be ubiquitous in protoplanetary disks and play a key role in planetary formation \citep{Boogert2015,Pontoppidan_2014, Musiolik_Wurm_2019, McClure_2023}. The shape and minimum position of ice bands can provide information on ice properties. However, investigating ice features is challenging and requires particular caution. Indeed, several studies have shown that these features depend on the system geometry as well as on the path of photons and the relative contributions of absorption and scattering \citep{Pontoppidan_2005, Sturm2023_HH48_ices, Dartois_2022, Martinien_2024, Martinien_2025}. Therefore, \citet{Martinien_2025} used model images to demonstrate that for sufficiently inclined disks 
ice features also depend on the location within the disk. Indeed, band shapes are completely different at the edges (scattering-dominated light) compared to the central point source (transmitted light). The James Webb Space Telescope (JWST) now enables new insights into the study of ices through spectrophotometry and will help to study ices behaviors in more details. 

However, the determination of ice abundance within the disk remains unclear. 
Indeed, the most common ice bands reach saturation in highly inclined disks due to radiative transfer effects, especially multiple scattering. This phenomenon of bands saturation has been observed in several models with integrated spectra: in the edge-on disk HH\,48\,NE \citep{Sturm2023_HH48_ices}, in the grazing-angle disk PDS\,453 \citep{Martinien_2024}, but also in generic models with spectra extracted at different locations within the disk \citep{Martinien_2025}. 

Fortunately, the geometry of highly inclined disks allows to probe their vertical structure and to measure the thickness of the disk. Observationally, these disks present a unique shape, with their two scattering nebulae separated by a central darklane. The thickness of the disk across the absorption band, measured as separation between the maximum of the two scattering nebulae, can be used to overcome the problem of band saturation. This separation is directly connected to the disk mass and dust opacity \citep{Watson_2004}. Therefore, measuring the disk thickness as a function of wavelength is a direct, quantitative probe of the opacity law of the solid particles in the disk. Thus, changes can appear in the distance between the two nebulae because of the addition of ice opacity over
continuum opacity. This can be directly linked to the abundance of ice within the disk and, when coupled with disk modeling, can serve as a tracer of ice abundance.

The best objects to study this are thus edge-on disks for which the two scattering nebulae are well separated. We studied two edge-on disks in Taurus (Tau\,042021 and HH\,30), and two in Ophiuchus  (Oph\,163131, and Flying Saucer); all are T Tauri stars. For the first three disks, JWST/NIRCam (Near Infrared Camera) images were analyzed by \citet{Duchene_2024}, \citet{Tazaki_2025} and \citet{Villenave_2024}, respectively, including the measurement of the distance between the two nebulae. The presence of multiple ice species has already been demonstrated in HH\,30, Tau\,042021 \citep{Pascucci_2025, Dartois_2025} and Flying Saucer \citep{Aikawa2012}.

In this paper, we first use synthetic model images to demonstrate that the presence of water ice in a disk can be inferred from the disk thickness and that the effect does not saturate even at high ice abundance. We then show that this phenomenon is observed in all four edge-on disks studied here. Finally, we propose a method to estimate the ice abundance in highly inclined disks, independently of spectroscopic analysis.

\section{Modeling setup and JWST data}
\label{sec:methods}

\begin{figure}[h!]
        \centering
        \includegraphics[width=0.41\textwidth]{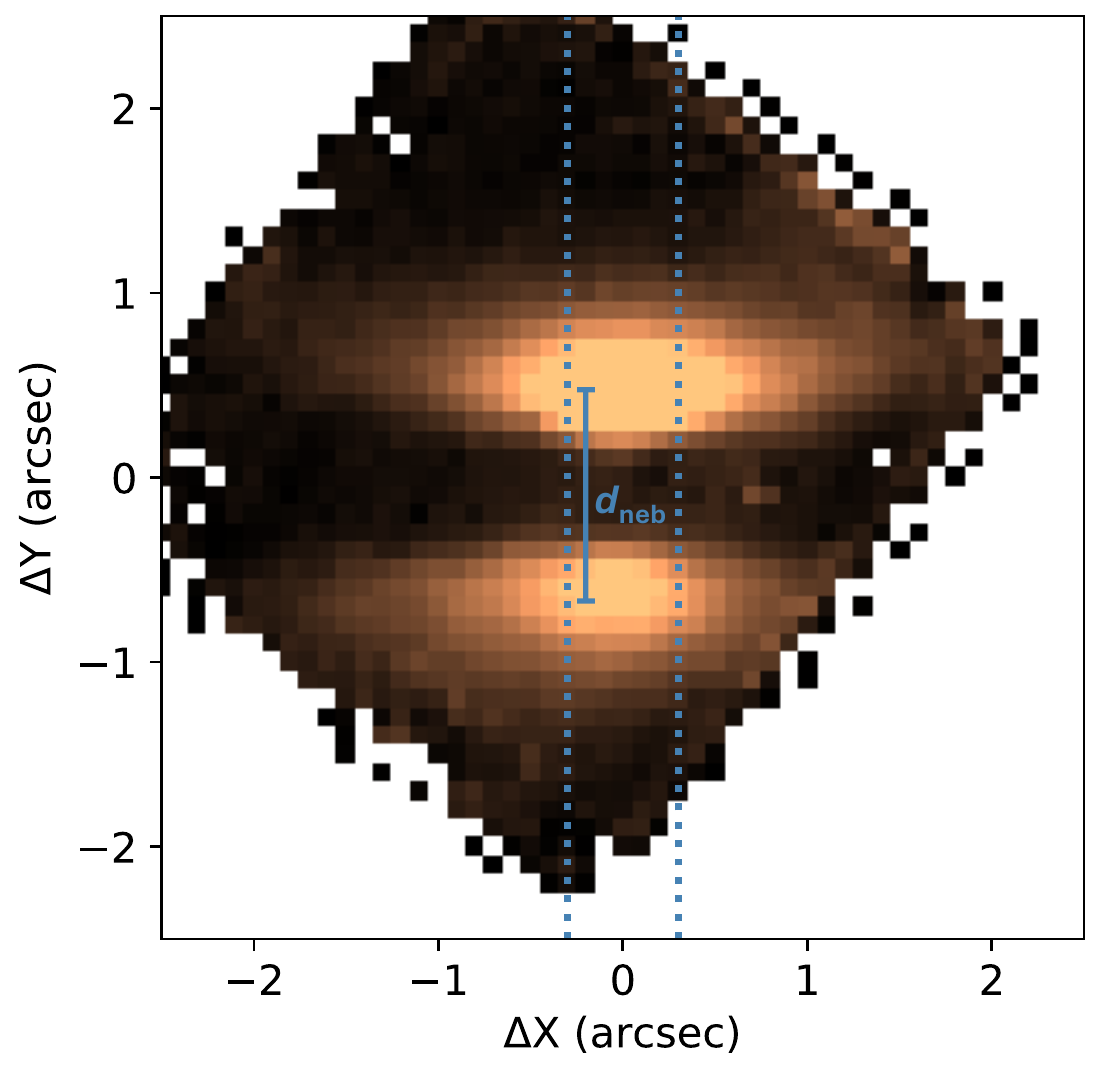}
        \caption{JWST/NIRSpec image of Tau\,042021 centered at 3.1\,\mic\, with a binning of 30 slices (0.01\,\mic\,). The solid blue line represents the distance between the two nebulae and the dotted blue lines represent the limits over which we measured this distance.}
        \label{fig:NIRSPEC}
    \end{figure}

We measured the disk thickness in models and observations of edge-on disks to monitor its variations as a function of wavelength. In this study we defined the disk thickness as the distance between \rev{the two scattering nebulae ($d_\mathrm{neb}$) at the position for which the darklane is narrowest}; see Fig. \ref{fig:NIRSPEC}. 

We studied the disk thickness between 2.0 and 4.0\,\mic\, in the edge-on disk model described in \citet{Martinien_2025} and composed of a mixture of 70$\%$ of astronomical silicates and 30$\%$ of water ice, in volume fraction.  We used the radiative transfer code MCFOST \citep{Pinte_2006} to vary the water-ice abundance in the disk by considering different volume fraction ranging from 0$\%$ to 90$\%$, \rev{sampled every 15$\%$}. 
Following the study of \citet{Martinien_2025}, we considered only water ice; however, we expect the same phenomenon for other ice species \rev{(CO$_{2}$ and CO) because the behavior of their opacity constants is similar to water ice.}. We applied a convolution by a wavelength-scaled Gaussian kernel as a simple approximation of the JWST point spread function.

We also used JWST/NIRSpec integral field unit 
archival data of Cycle 1 (program number: 1621, PI: I. Pascucci) and Cycle 3 (program number: 5299, PI: J. Bergner) with calibration software version 1.18.0 to study the disk thickness in observations. In most cases, the two surface nebulae are well separated. However, Oph\,163131 is much thinner and the distance between the nebulae is more difficult to estimate, especially at the longest wavelengths. We used three filters with three grisms: G140H, G235H, G395H for HH\,30, and Tau\,042021 and two filters with two grisms: G235H, G395H for Flying Saucer and Oph\,163131, providing a wavelength coverage between 0.97 and 5.27\,\mic\, and between 1.66 and 5.27\,\mic\,, respectively. We removed the bad slices for which the disk was cut and we also applied a binning of 20 slices except for Tau\,042021 for which we binned over 30 slices along the spectral dimension to reduce the noise. Despite the NIRSpec pixel scale of 0.1\arcsec, the achieved SNR is sufficient to allow a precise measurement of the distance between the two nebulae. \rev{This results in an effective sampling that is about 10 times lower than the instrument resolution, ensuring that all points are independent.}

We first extracted the vertical profile at the central pixel of the disk for the model images as well as observations of HH\,30 and Oph\,163131. However, for Tau\,042021 and Flying Saucer with a lower signal-to-noise ratio,  we integrated over 0.6$\arcsec$  and 0.4$\arcsec$ respectively around the central pixel. 
We then used Gaussian fits to the brightness profile to find the position of the peaks of each nebula. We note that the brightness profile is not exactly Gaussian and using different peak-finding methods can yield slightly different disk thicknesses, but the general trend and spectral features remain unchanged. In most cases, we performed the fit one nebula at a time as they are well separated in the images.
However, for the thin disk Oph\,163131, the lack of clear separation between the two nebulae required the use of a simultaneous double Gaussian fit, 
as no distinct second peak could be identified. Even then, the results have to be considered with caution at longer wavelengths. Indeed, beyond 4\,\mic\,, the separation is so thin that the Gaussian fit is biased and the estimated distance appears to rise, in contradiction with the results from the better-sampled NIRCam images \citep{Villenave_2024}.

\section{Results}
\label{sect:results}

\subsection{Disk thickness in models}

\begin{figure*}[htb]
        \centering
        \includegraphics[width=0.46\textwidth]{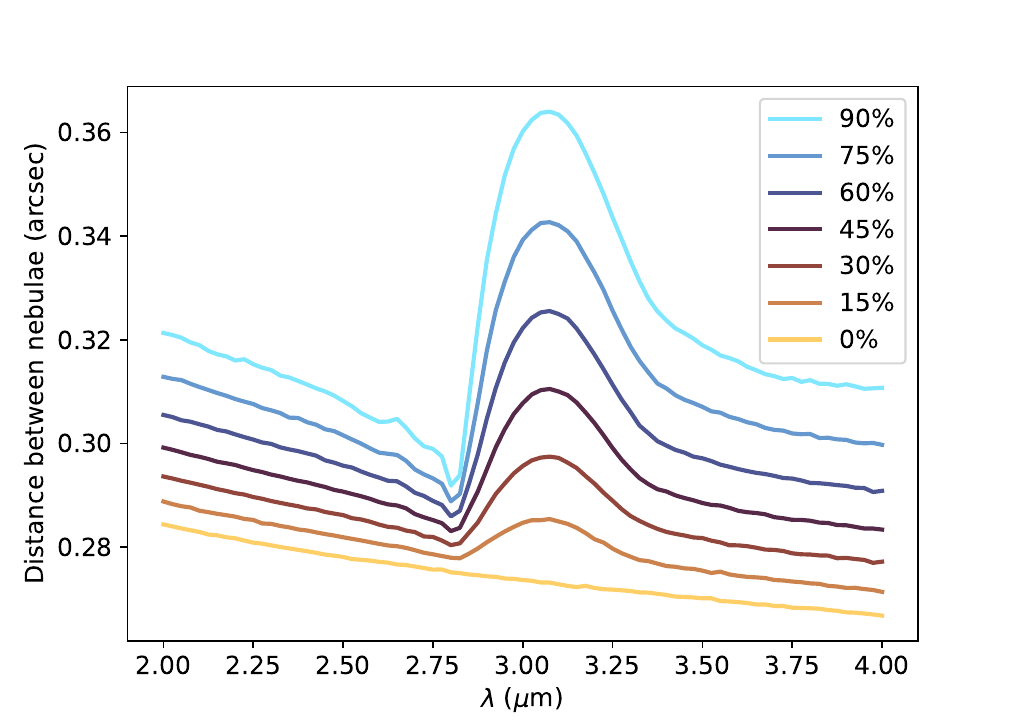}
        \includegraphics[width=0.443\textwidth]{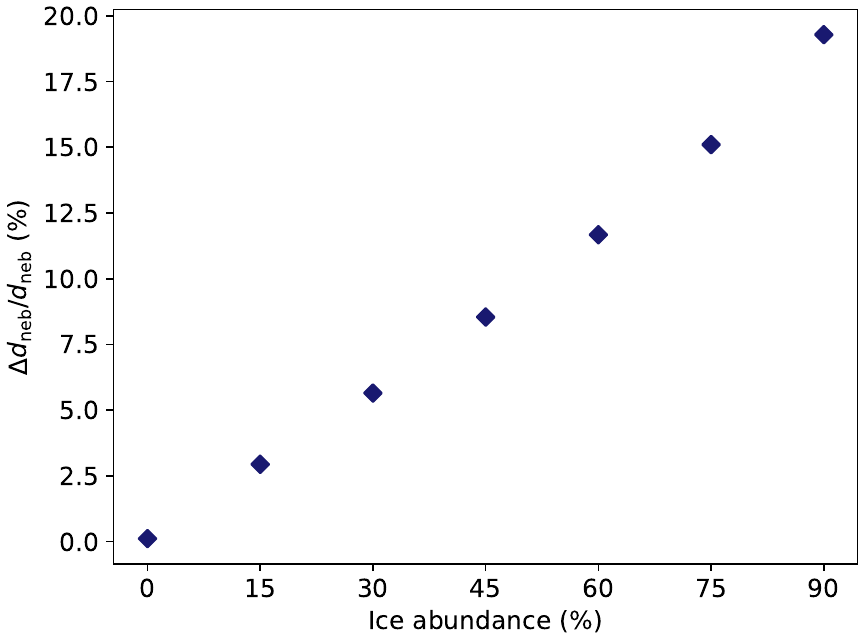}
        \caption{Left: Distance between the two nebulae as a function of wavelength in the edge-on disk model presented in \citet{Martinien_2025} for different water ice abundances. \rev{Right: Bumps heights relative to the continuum in the same models as a function of ice abundance.}}
        \label{fig:MODELS}
    \end{figure*}

Fig. \ref{fig:MODELS} (left panel) shows the distance between the two nebulae in the edge-on disk model at 90\degr\, in \citet{Martinien_2025} for different water ice abundances. For each abundance, 
$d_\mathrm{neb}$ decreases with wavelength because of the decreasing continuum dust opacity. However, $d_\mathrm{neb}$ increases with a feature (or bump) around $\sim$3.1\,\mic\ which corresponds to the position of the water-ice band. We defined the height of the bump relative to the continuum, $\Delta$$d_\mathrm{neb}$/$d_\mathrm{neb}$, with $d_\mathrm{neb}$ determined at the wavelength of the peak. \rev{We can clearly see that this height is correlated with the ice abundance and is not subject to saturation as demonstrated by the nearly linear relationship shown in the right panel of Fig. \ref{fig:MODELS}}. Indeed, when there is no water ice in the model, the distance monotonically decreases with wavelength without any bump. However, an increase in the abundance of ice leads to a more prominent change in thickness across the 3\,\mic\ ice band. 

The shape of the bump also changes with the ice abundance with a more roundish shape for low abundance and a sharper bump for high abundance. The continuum on both sides of the water-ice feature, especially at longer wavelengths, is higher with the addition of water ice. This results from an increase of the extinction and scattering opacity at higher ice abundances, as ice-dominated grains scatter more efficiently.
\rev{We emphasize that the proportionality between ice abundance and $\Delta$$d_\mathrm{neb}$/$d_\mathrm{neb}$ is model-dependent and will vary with parameters such as system inclination, disk mass, and scale height. In particular, a larger scale height will amplify the increase in disk thickness with ice abundance, while a smaller scale height will diminish this effect.}

\subsection{Disk thickness in observations}
\begin{figure*}[h]
        \centering
        \includegraphics[width=0.85\textwidth]{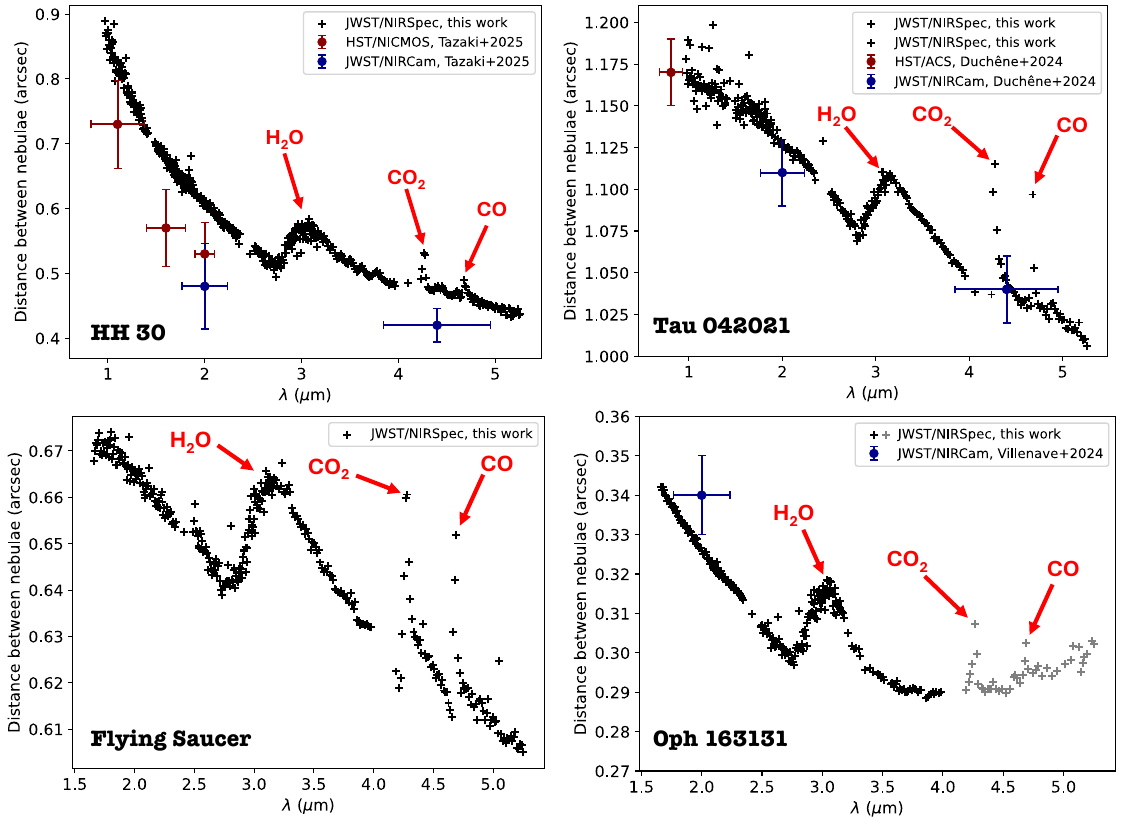}
        \caption{Distance between the two nebulae as a function of wavelength in JWST/NIRSpec observations. In Oph\,163131, the gray points correspond to a wavelength range where the two nebulae are no longer clearly separated and the results of the gaussian fit are likely biased upwards. The horizontal bars correspond to the wavelength coverage of each imaging filter.}
        \label{fig:OBSERVATIONS}
    \end{figure*}

The same phenomenon is observed in JWST/NIRSpec observations of HH\,30, Tau\,042021, Flying Saucer and Oph\,163131 (Fig. \ref{fig:OBSERVATIONS}). Indeed, $d_\mathrm{neb}$ decreases with wavelength due to reduced dust opacity, except around $\sim$3.1\,\mic\ where the $d_\mathrm{neb}$ increases due to the addition of opacity from water ice. 
The shift is subtle in the images, about half a pixel, but it is easily measurable given the high SNR of the data, as illustrated by the smoothness of the resulting plots in Fig \ref{fig:OBSERVATIONS}. \rev{The resulting errors can be determined as the scatter along the curves and are the product of the SNR and the thickness of the darklane and thus, depend on each object.} We note that the offset between the absolute values of our distance measurements and those from JWST/NIRCam archival data presented in Fig. \ref{fig:OBSERVATIONS} arises from a combination of the difference in spatial sampling between NIRCam and NIRSpec and the use of different measurement methods. Crucially, these effects do not affect the overall trend. 

Furthermore, there are also bumps for all disks around $\sim$4.2 and $\sim$4.6\,\mic\ corresponding to CO$_{2}$ and CO ices, respectively. To go beyond spectroscopic analyses, our study unambiguously demonstrates that the ices are located inside the disks themselves and not from the foreground absorption as the latter would not affect the apparent disk morphology.

Moreover, the ratios of $\Delta$$d_\mathrm{neb}$/$d_\mathrm{neb}$ for H$_{2}$O, CO$_{2}$ and CO ices differ among the four disks, suggesting different relative abundances between these ices species. 
Finally, we note that the amplitude of the bumps is not correlated with the depth of the same ice bands measured in spectra presented in \citet{Aikawa2012} and \citet{Pascucci_2025}.

\section{Discussion and conclusions}

The bumps present in models and observations are caused by an increase of opacity due to the presence of ices 
on top of the continuum opacity, raising the height of the scattering surface at the center of the bands. The height of bumps is clearly correlated with the ice abundance within the disks (Fig. \ref{fig:MODELS}). However, the ice location and abundance depend on the temperature structure as well as the chemistry \citep{Woitke_2022}. Therefore, we refrain from attributing an ice abundance to a bump height, or to compare ice abundance between disks. Within a given disk, if we assume that the ratio of bump heights traces the abundance ratio, then the different $\Delta$$d_\mathrm{neb}$/$d_\mathrm{neb}$ measured in the various disks suggest variations in relative species abundance.
According to this, it will be possible to speculate 
about relative ice abundances between H$_{2}$O, CO$_{2}$ and CO within each disks by looking at bumps heights relative to the continuum, the values for which we estimated the continuum by linear interpolation from either side of the band are listed in Table \ref{tab:bumps}. 
There is a variety of ice composition in the disk sample. HH\,30 is dominated by the H$_2$O feature, Tau\,042021 exhibits stronger CO$_2$ and CO signatures, whereas Flying Saucer and Oph\,163131 display more balanced contributions from the three ice species. 
These differences can be driven by sublimation effects due to differences in disk size, star properties or environment, and therefore, quantitative ice absolute abundance estimates will require physico-chemical models tailored to each sources.

\setlength{\tabcolsep}{4pt}
\begin{table}[h!]
	\centering
 \caption{Bumps heights relative to continuum ($\Delta$$d_\mathrm{neb}$/$d_\mathrm{neb}$) of the three ice species for the four edge-on disks.
 } 
        \footnotesize 
        \begin{threeparttable}
\begin{tabular}{cccccccc}
		 \hline
          \hline
		&  \multicolumn{3}{c} {$\Delta$$d_\mathrm{neb}$/$d_\mathrm{neb}$($\%$)} & & \multicolumn{3}{c} {$\Delta$$d_\mathrm{neb}$/$d_\mathrm{neb}$($\%$)} \\
               &  H$_{2}$O & CO$_{2}$ & CO &  &H$_{2}$O & CO$_{2}$ & CO\\
		 \hline
		HH\,30 & $\sim$14 & $\sim$12 & $\sim$6 & Flying Saucer & $\sim$4 & $\sim$5 & $\sim$5 \\
        Tau\,042021  & $\sim$3 & $\sim$7 & $\sim$6 & Oph\,163131 &  $\sim$7 & $\sim$6 & $\sim$4\\

        \hline
	\end{tabular}
    \tablefoot{The values for CO$_{2}$ and CO in Oph\,163131 are likely biased (see Sect. \ref{sec:methods}).}
    \end{threeparttable}
      \label{tab:bumps}
\end{table}

There are also differences in bump widths between different ice species. 
This is linked to the difference of bands width in spectra. Indeed, CO$_{2}$ and CO bumps are narrower than H$_{2}$O bump. This can be explained by the wavelength coverage of H$_{2}$O ice band which is wider \citep{Gibb_2004} than the wavelength coverage of CO$_{2}$ and CO ice bands because of their refractive index.
Moreover, as showed in Sect. \ref{sect:results}, the shape of the water-ice bump is more roundish for low abundance and sharper for high abundance (Fig. \ref{fig:MODELS}). Hence, the bump shape itself may provide additional insight into ice abundances.

In more details, the method consists of building a model for which the disk thickness, and thus the opacities, is in agreement with the observed thickness and then considering the ice abundance as the only tunable parameter \rev{to match the observed bump height}. The method is not restricted to edge-on disks, but can also be applied to less inclined disks such as those seen at grazing angle, as long as the geometry allows both scattering nebulae to be observed. For instance, we applied the same method in our model with a grazing-angle inclination and the same trend is observed.

Thanks to the measure of the disk thickness in highly inclined disks, it will thus be possible to determine the abundance of ice by fitting specific disk models. The only caveat is that, in order to obtain reliable measurements at all wavelengths, the distance between the two nebulae must be visible, which requires the disk \rev{surfaces to be well-separated}. This limitation is well exemplified by the case of Oph\,163131. As our models demonstrated, this method is not affected by the saturation problem inherent to spectroscopic studies, and it thus promises to be more robust and broadly applicable.

\bibliography{BibLau.bib}   

\begin{acknowledgements}
This project has received funding from the European Research Council (ERC) under the European Union’s Horizon Europe research and innovation program (grant agreement No. 101053020, project Dust2Planets, PI: F. M\'enard).
E.D. and J.A.N. acknowledge support from the Thematic Action ‘Physique et Chimie du Milieu Interstellaire’ (PCMI) of INSU Programme National ‘Astro’, with contributions from CNRS Physique $\&$ CNRS Chimie, CEA, and CNES.
The data presented in this paper were obtained from the Mikulski Archive for Space Telescopes (MAST) at the Space Telescope Science Institute. The  JWST observations analyzed can be accessed via \url{https://doi.org/10.17909/4tha-mz43}.
\end{acknowledgements}

%
%

\label{LastPage}

\end{document}